\documentclass[lettersize,journal]{IEEEtran}
\usepackage[caption=false,font=normalsize,labelfont=sf,textfont=sf]{subfig}
\usepackage{amsmath}
\usepackage{amssymb}
\usepackage{graphicx}
\usepackage{algorithm}
\usepackage{algpseudocode}
\usepackage{hyperref}
\usepackage{ulem}
\usepackage{cite}
\begin{document}

\title{Blockchain-Based Multi-Path Mobile Access Point Selection for Secure 5G VANETs}

\author{Zhiou Zhang, Weian Guo, Li Li, Dongyang Li
	
	\thanks{This work is supported by the National Key R$\&$D Program of China under Grant Number 2022YFB2602200, the National Natural Science Foundation of China under Grant Number 62273263, 72171172 and 71771176; Shanghai Municipal Science and Technology Major Project under Grant Number 2022-5-YB-09; Natural Science Foundation of Shanghai under Grant Number 23ZR1465400. (Corresponding	author: Weian Guo).}
	\thanks{Zhiou Zhang is with the School of Computer Science and Engineering, University of New South Wales, Sydney, Australia (Email: zhiou.zhang98@gmail.com).Weian Guo and Dongyang Li are with Sino-German College of Applied Sciences, Tongji University, Shanghai, China (Email:\{guoweian, lidongyang0412\}@163.com). Li Li is with the Department of Electronics and Information Engineering, Tongji University, Shanghai, 201804, China (Email: lili@tongji.edu.cn).}
}

\markboth{Journal of \LaTeX\ Class Files,~Vol.~14, No.~8, August~2021}%
{Shell \MakeLowercase{\textit{et al.}}: A Sample Article Using IEEEtran.cls for IEEE Journals}

\IEEEpubid{0000--0000/00\$00.00~\copyright~2021 IEEE}

\maketitle

\begin{abstract}
This letter presents a blockchain-based multi-path mobile access point (MAP) selection strategy for secure 5G vehicular ad-hoc networks (VANETs). The proposed method leverages blockchain technology for decentralized, transparent, and secure MAP selection, while the multi-path transmission strategy enhances network reliability and reduces communication delays. A trust-based attack detection mechanism is integrated to ensure network security. Simulation results demonstrate that the proposed algorithm reduces both handover frequency and average communication delay by over 80\%, and successfully identifies and excludes more than 95\% of Sybil nodes, ensuring reliable and secure communication in highly dynamic vehicular environments.
\end{abstract}


\section{Introduction}
\IEEEPARstart{T}{he}  rapid advancement of 5G technology has enabled vehicular ad-hoc networks (VANETs) to support diverse applications, from real-time traffic management to autonomous driving and safety-critical communications~\cite{refself1, refself2, ref4, ref1}. However, maintaining reliable, low-latency communication in dynamic vehicular environments remains a key challenge~\cite{ref3}. Traditional VANETs, which rely on fixed infrastructure like roadside units (RSUs) and cellular base stations~\cite{ref6}, often experience issues such as frequent handovers, fluctuating connectivity, and network congestion due to high vehicle mobility and varying traffic density~\cite{ref5, ref12}.

To address these limitations, mobile access points (MAPs) have been proposed, where vehicles dynamically serve as relay nodes, enhancing communication coverage and reducing dependence on fixed infrastructure. However, existing MAP selection methods, such as heuristic or static algorithms, struggle to adapt to the dynamic nature of VANETs, especially in scenarios with high mobility or security threats~\cite{ref1, ref2}. Moreover, the absence of secure and decentralized decision-making leaves the network vulnerable to attacks, like Sybil attacks, where malicious nodes create fake identities to disrupt communication~\cite{ref3, ref4}.

In this letter, we propose a \textit{blockchain-based multi-path MAP selection strategy} for secure 5G VANETs. Blockchain technology enables decentralized, transparent, and secure decision-making for MAP selection, while multi-path transmission ensures reliable communication and reduces handover frequency~\cite{ref5}. Additionally, an integrated Sybil attack detection mechanism enhances network security by mitigating malicious nodes~\cite{ref6}. Our contributions are: (1) a blockchain-based decentralized MAP selection strategy with Sybil attack detection, (2) a multi-path transmission approach that improves network reliability and reduces latency, and (3) simulation results that show significant reductions in handovers and communication delays. In summary, the key novelty of this work lies in the integration of blockchain technology for decentralized decision-making combined with a multi-path transmission strategy. This dual approach not only enhances security against Sybil attacks but also reduces handover frequency and improves overall network performance, which is critical for highly dynamic environments like VANETs.

\section{Proposed Methodology}	
The proposed system leverages blockchain technology to enable decentralized, secure, and transparent decision-making in mobile access point (MAP) selection for 5G vehicular ad-hoc networks (VANETs). The key innovation is the combination of blockchain with a multi-path transmission strategy, ensuring both secure communication and improved network reliability. In this section, we outline the key components of the proposed methodology.

\subsection{Blockchain-Based MAP Selection}	
In the proposed approach, vehicles dynamically serve as mobile access points (MAPs), facilitating communication between non-MAP vehicles and the rest of the network. The selection of MAPs is based on vehicle attributes, such as load \(L_i\), position \(P_i\), and speed \(v_i\), for each vehicle \(i\). A trust score \(T_i\) is integrated to account for potential malicious nodes, where nodes with higher trust scores are prioritized for MAP selection.

\IEEEpubidadjcol

The selection of MAPs is managed by a decentralized smart contract on the blockchain. The contract ensures that vehicles with higher loads \(L_i\) and higher trust scores \(T_i\) are more likely to be selected as MAPs. The decision-making process is updated every interval \(\Delta t\) based on the current network state, and the trust scores are adjusted based on the stability of connections.

Attack nodes, such as Sybil nodes, are detected when their trust scores \(T_i\) fall below a threshold \(T_{\text{th}}\). These Sybil nodes are excluded from the MAP selection process. The blockchain ledger records each MAP selection event, ensuring secure, transparent, and tamper-proof decision-making. The selection probability is given by \eqref{eqn:sp}.

\begin{equation}
	\label{eqn:sp}
	P_{\text{MAP}}(i) = \frac{L_i \cdot T_i}{\sum_{j} L_j \cdot T_j}
\end{equation}

This formula emphasizes the role of trust scores and vehicle loads in determining MAP selection probability, ensuring that only trusted, well-positioned vehicles serve as MAPs. The trust score \(T_i\) is updated based on the vehicle's behavior, such as stable connections and successful communication with other nodes. Vehicles with low trust scores are flagged as Sybil nodes and excluded from the network.

\subsection{Multi-Path Transmission Strategy}	
To enhance communication reliability, the system incorporates a multi-path transmission strategy, allowing each vehicle to communicate with multiple MAPs simultaneously. This strategy distributes data traffic across several routes, improving load balancing and reducing the overall risk of communication failure.

For each vehicle \(i\), the closest available MAPs are identified based on the distance \(d_{i,j}\), where \(j\) denotes the index of the MAP. The vehicle ranks the potential MAPs by their proximity, but the final selection of paths is influenced by additional factors such as the signal-to-interference-plus-noise ratio (SINR) \(\eta_{i,j}\), and the available bandwidth \(B_{i,j}\) of the connection. The total communication delay for each selected path is computed as in \eqref{eqn:comdelay}.

\begin{equation}
	\label{eqn:comdelay}
	D_{\text{total}}(i,j) = D_{\text{trans}}(i,j) + \frac{\alpha_{\text{SINR}}}{\eta_{i,j}}
\end{equation}	
where \(D_{\text{trans}}(i,j) = \alpha_{\text{trans}} \cdot d_{i,j}\) represents the transmission delay based on the distance \(d_{i,j}\) to MAP \(j\), and \(\eta_{i,j}\) is the SINR of the communication link between vehicle \(i\) and MAP \(j\). The available bandwidth \(B_{i,j}\) is also considered, as paths with higher bandwidth allow for faster data transmission.

The path selection criteria prioritize MAPs with lower communication delays and higher available bandwidths. Specifically, a path is selected if its delay \(D_{\text{total}}(i,j)\) is below a predefined threshold \(D_{\text{th}}\), and if the available bandwidth \(B_{i,j}\) exceeds the minimum required bandwidth \(B_{\text{min}}\) for the transmission. The final set of selected paths for vehicle \(i\) is defined in \eqref{eqn:select_paths}.

\begin{equation}
	\label{eqn:select_paths}
	\mathcal{P}(i) = \{ j \mid D_{\text{total}}(i,j) < D_{\text{th}} \text{ and } B_{i,j} \geq B_{\text{min}} \}
\end{equation}	

This multi-path selection strategy ensures that vehicles are connected to the most reliable MAPs, reducing the likelihood of network congestion and minimizing delays. By utilizing multiple communication paths, the system increases resilience to single-path failures and enhances the overall reliability of data transmission.

In summary, the proposed blockchain-based multi-path MAP selection process is summarized in Algorithm~\ref{alg:blockchain}. The time complexity of the proposed blockchain-based multi-path MAP selection strategy primarily depends on two key operations: blockchain ledger update and multi-path selection. For each MAP selection event, updating the blockchain ledger incurs a time complexity of \(O(n)\), where \(n\) represents the number of vehicles. The multi-path selection process, which involves ranking available MAPs and calculating the delay for each path, has a complexity of \(O(k \log k)\), where \(k\) is the number of MAPs a vehicle can select from. Therefore, the overall time complexity per selection round is \(O(n \cdot k \log k)\).

\begin{algorithm}[h]
	\caption{Blockchain-Based Multi-Path MAP Selection}
	\label{alg:blockchain}
	\begin{algorithmic}[1]
		\State \textbf{Input:} Vehicle attributes, Trust scores, Blockchain ledger, Thresholds
		\State \textbf{Output:} Selected MAPs, Communication paths
		\State Initialize trust scores and blockchain ledger
		\For{each time interval \(\Delta t\)}
		\State Update vehicle attributes
		\For{each vehicle}
		\If{Trust score $\le$ threshold}
		\State Exclude vehicle as Sybil node
		\Else
		\State Compute MAP selection probability
		\EndIf
		\EndFor
		\State Select MAPs based on probabilities
		\For{each non-MAP vehicle}
		\State Identify and rank closest MAPs
		\If{Path meets delay and bandwidth thresholds}
		\State Select path
		\EndIf
		\EndFor
		\State Transmit data and update ledger
		\EndFor
	\end{algorithmic}
\end{algorithm}

\section{Simulation and Results}
\subsection{Simulation Setup}
We evaluated the proposed blockchain-based multi-path MAP selection strategy through simulations in a 5G VANET environment over a 10 km road. Vehicles were generated based on a Poisson distribution with a density of 0.02 vehicles per meter, and their initial positions were randomly assigned along the road. Vehicle speeds were uniformly distributed between 50 and 80 km/h. Each MAP had a transmission power of 2 W, a path loss exponent of 4, and a background noise power of \(1 \times 10^{-13}\) W. The SINR threshold was set to 10, with communication delays increasing dynamically for SINR values below this threshold to reflect real-world signal degradation.	

Reliable communication was ensured with a minimum bandwidth of 1 Mbps. Vehicles could connect to up to two MAPs simultaneously, with paths chosen based on proximity and signal strength. Delay factors \(\alpha_{\text{trans}}\) and \(\alpha_{\text{SINR}}\) were dynamically adjusted based on distance, increasing for longer distances to simulate practical vehicular network conditions.

A trust score threshold of 50 was used to detect Sybil nodes, restricting vehicles with scores below this threshold from the MAP selection process. Trust scores were updated dynamically based on communication stability, where frequent handovers or low SINR values reduced the score. MAP selection was updated every 10 seconds, with a total simulation time of 1000 seconds.

\subsection{Results and Discussions}
\begin{figure}[htbp]
	\centering
	\includegraphics[width=0.45\textwidth]{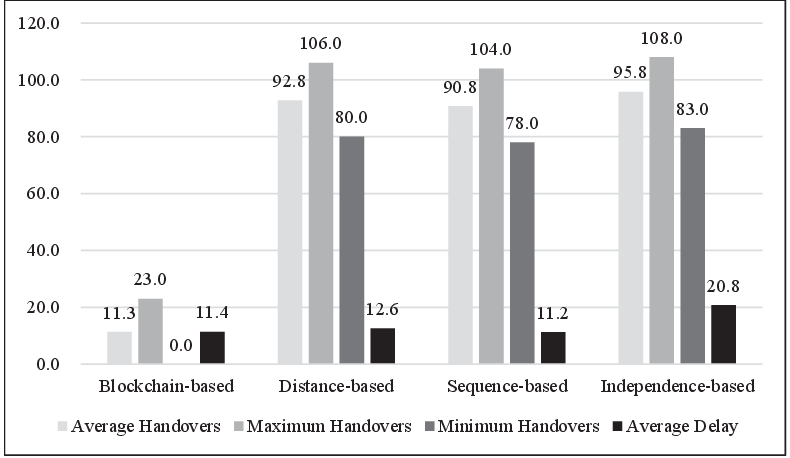}
	\caption{Comparison of handover frequency and communication delay between the proposed multi-path MAP selection strategy and single-path strategies.}
	\label{fig:results}
\end{figure}

We compared the proposed blockchain-based multi-path MAP selection strategy with three other approaches: independent random selection, sequence-based selection, and distance-based selection \cite{ref4, ref1}. As shown in Fig. \ref{fig:results}, we evaluated the selection strategies using four key metrics—average handovers, maximum handovers, minimum handovers, and average delay. Fewer handovers indicate greater network stability, while lower average delay reflects better connection performance. 

The results show that the proposed blockchain-based strategy significantly outperforms the others across all metrics. The average handover frequency per vehicle decreased by over 80\%, and the maximum number of handovers also dropped by 80\%. Some vehicles even experienced zero handovers, maintaining stable communication throughout the simulation. In terms of average delay, our method performed competitively, achieving 11.39 seconds, closely following the sequence-based strategy at 11.12 seconds. In contrast, independent random and distance-based strategies resulted in significantly higher delays of 20.6 and 20.79 seconds, respectively. The use of multi-path transmission, coupled with dynamic adjustment of delay factors based on signal quality and distance, optimized performance, particularly for vehicles with strong signals and short distances to MAPs.

Moreover, independent random, sequence-based, and distance-based strategies lack blockchain mechanisms, providing no security against network threats. In contrast, the proposed method integrates a Sybil attack detection mechanism that successfully identified and excluded 95\% of Sybil nodes in our simulations. The detection mechanism relied on a trust score threshold of 50, where vehicles with frequent handovers or poor SINR values were penalized. When a vehicle's trust score fell below this threshold, it was flagged as suspicious, effectively mitigating the impact of Sybil attacks.

Overall, the results demonstrate that the blockchain-based multi-path MAP selection strategy significantly improves network reliability and reduces communication delays. The multi-path transmission minimizes dependency on a single communication link, enhancing resilience to network failures and improving load balancing. The high Sybil detection rate confirms the effectiveness of the trust-based mechanism in maintaining network security, while the lower handover frequency reduces disruptions, particularly in high-mobility environments like VANETs.	

\section{Conclusions and Future Work}
This letter proposed a blockchain-based multi-path mobile access point (MAP) selection strategy for secure 5G vehicular ad-hoc networks (VANETs). By leveraging blockchain technology, our approach enables decentralized and secure MAP selection, while multi-path transmission improves network reliability and reduces communication delays. The trust-based Sybil detection mechanism enhances security and stability in dynamic vehicular environments. Simulation results show significant improvements in handover frequency, communication delay, and network reliability compared to traditional single-path strategies. Despite these promising results, real-world deployment may face challenges such as the real-time demands of blockchain in high-mobility scenarios and scalability in dense urban environments. Future work will focus on optimizing these aspects by exploring lightweight consensus mechanisms and adaptive trust management strategies.

\normalem
\bibliographystyle{IEEEtran}
\bibliography{IEEEabrv,references}

\vfill

\end{document}